\documentclass[11pt]{cernrep}
\usepackage{graphicx}
\def\lsim{\raise0.3ex\hbox{$<$\kern-0.75em\raise-1.1ex\hbox{$\sim$}}}
\def\gsim{\raise0.3ex\hbox{$>$\kern-0.75em\raise-1.1ex\hbox{$\sim$}}}

\begin{document}

\title{THE DENSITY OF STATES METHOD AT FINITE CHEMICAL POTENTIAL\\[-2.5cm]
\normalsize \tt \hfill BNL-NT-05/51\\[2.5cm]}

\author{Christian Schmidt\footnote{$\;\;$Speaker}$\;^{1}$, Zoltan Fodor$^{2,3}$ and Sandor Katz$^{3}$}

\institute{
$^1$ Physics Department, Brookhaven National Laboratory, Bldg. 510A, Upton, NY, 11973, USA\\
$^2$ Department of Physics, University of Wuppertal, Gaussstrasse 20, D-42119, Germany\\
$^3$ Institute for Theoretical Physics, E\"otv\"os University, Pazmany 1, H-1117 Budapest,
Hungary
}

\maketitle

\begin{abstract}
We study the density of states method to explore the phase diagram
of the chiral transition on the temperature
and quark chemical potential plane. Four quark flavors are used in the
analysis. Though the method is quite expensive small lattices
show an indication for a triple-point connecting three different phases on
the phase diagram.
\end{abstract}

\section{Introduction}
To clarify the phase diagram of QCD and thus the nature of matter under
extreme conditions is one of the most interesting and fundamental tasks of
high energy physics.
Lattice QCD has been shown to provide important and reliable information from
first principals on QCD at zero density. However, Lattice QCD at finite
densities has been harmed by the complex action problem ever since its
inception. For $\mu>0$ the determinant of the fermion
matrix ($\rm{det}M$) becomes complex. Standard Monte Carlo techniques using
importance sampling are thus no longer applicable when calculating observables
in the grand canonical ensemble according to the partition function
\begin{equation}
Z_{GC}(\mu)=\int \mathcal{D}U\; \rm{det}M[U](\mu) \exp\{-S_G[U]\}.
\label{eq:Z_GC}
\end{equation}
Recently many different methods have been developed to circumvent the complex
action problem for small $\mu/T$ \cite{Fodor:2001au, methods}. For a recent 
overview see also \cite{overview}.

\section{Formulation of the method\label{sec:method}}
A very general formulation of the DOS
method is the following: One exposed parameter ($\phi$) is fixed. The
expectation value of a thermodynamic observable ($O$), according to the usual
grand canonical partition function (\ref{eq:Z_GC}), can be recovered by the
integral
\begin{equation}
<O>=\int d\phi \, \left<Of(U)\right>_\phi \rho(\phi)
\left/ \int d\phi \, \left<f(U)\right>_\phi \rho(\phi)\right.
\label{eq:dos_obs}
\end{equation}
where the density of states ($\rho$) is given by the constrained partition
function:
\begin{equation}
\rho(x)\equiv Z_\phi(x)=\int \mathcal{D}U\, g(U) \, \delta( \phi - x ).
\label{eq:dos_Z}
\end{equation}
With $\left<~\right>_\phi$ we denote the expectation value with respect to the
constrained partition function. In addition, the product of the weight
functions $f,g$ has to give the correct measure of $Z_{GC}$:
$fg=\rm{det}M\exp\{-S_G\}$. This idea of reordering the partition
functions is rather old and was used in many different cases
\cite{dos, LUO, Ambjorn}
The advantages of this additional integration becomes
clear, when choosing $\phi=P$ and $g(U)=1$. In this case $\rho(\phi)$ is
independent of all simulation parameters. The observable can be calculated as
a function of all values of the lattice coupling $\beta$. If one has stored
all eigenvalues of the fermion matrix for all configurations, the observable
can also be calculated as a function of quark mass ($m$) and number of
flavors\cite{LUO} ($N_f$). In this work we chose
\begin{equation}
\phi=P \qquad \mbox{and} \qquad
g=\left|{\rm det}M \right| \exp\{-S_G\}, \qquad f=\exp\{i\theta\}.
\label{eq:con}
\end{equation}
In other words we constrain the plaquette and perform simulations with measure
$g$. In practice, we replace the delta function in Equation~(\ref{eq:dos_Z}) by a
sharply peaked potential \cite{Ambjorn}. The constrained partition function
for fixed values of the plaquette expectation value can then be written as
\begin{equation}
\rho(x) \approx \int {\cal D}U\; g(U) \exp\left\{- V(x)\right\},
\end{equation}
where $\exp\{-V(x)\}$ is a Gaussian potential with
\begin{equation}
V(x)=\frac{1}{2}\gamma\left(x-P\right)^2.
\end{equation}
We obtain the density of states ($\rho(x)$) by the fluctuations of the actual
plaquette $P$ around the constraint value $x$. The fluctuation dissipation
theorem gives
\begin{equation}
\frac{d}{dx}\ln \rho(x)=<x-P>_x.
\end{equation}
Before performing
the integrals in Equation~(\ref{eq:dos_obs}) we compute from an ensemble 
generated at $(\mu_0,\beta_0)$:
\begin{eqnarray}
\label{eq:rew1}
\left<Of(U)\right>_x(\mu,\beta)
&=&\left<Of(U)R(\mu,\mu_0,\beta,\beta_0)\right>_x
/\left<R(\mu,\mu_0,\beta,\beta_0)\right>_x,\\
\label{eq:rew2}
\left<f(U)\right>_x(\mu,\beta)
&=&\left<f(U)R(\mu,\mu_0,\beta,\beta_0)\right>_x
/\left<R(\mu,\mu_0,\beta,\beta_0)\right>_x,\\
\label{eq:rew3}
\frac{d}{dx}\ln\rho(x,\mu,\beta)
&=&\left<(x-P)R(\mu,\mu_0,\beta,\beta_0)\right>_x.
\end{eqnarray}
Here $R$ is given by the quotient of the measure $g$
at the point $(\mu,\beta)$ and at the simulation point $(\mu_0,\beta_0)$,
\begin{equation}
R(\mu,\mu_0,\beta,\beta_0)=g(\mu,\beta)/g(\mu_0,\beta_0)
=\frac{|{\rm det}(\mu)|}{|{\rm det}(\mu_0)|}\exp\{S_G(\beta)-S_G(\beta_0)\}.
\end{equation}
Having calculated the expressions~(\ref{eq:rew1})-(\ref{eq:rew3}), we are
able to extrapolate the expectation value of the observable~(\ref{eq:dos_obs})
to any point $(\mu,\beta)$ in a small region around the simulation point
$(\mu_0,\beta_0)$. For any evaluation of $\left<O\right>(\mu,\beta)$, we
numerically perform the integrals in Equation~(\ref{eq:dos_obs}). We also
combine the data from several simulation points to interpolate between them.

\section{Simulations with constrained Plaquette}
The value we want to constrain is the expectation value of the global
plaquette, which is given on every gauge configuration by the sum over all
lattice points ($y$) and directions ($\mu\nu$) of the local plaquette
$P_{\mu\nu}(y)$ and its adjoint $P^\dagger_{\mu\nu}(y)$,
\begin{equation}
P=\sum_y\sum_{1\le\mu<\nu\le4} \frac{1}{6}\left[
{\rm Tr}P_{\mu\nu}(y) + {\rm Tr}P^\dagger_{\mu\nu}(y) \right].
\end{equation}
Since the plaquette is also the main part of the gauge action,
\begin{equation}
S_G=-\beta\sum_x\sum_{1\le\mu<\nu\le4}\left\{\frac{1}{6}\left[
    {\rm Tr}P_{\mu\nu}(x) + {\rm Tr}P^\dagger_{\mu\nu}(x)\right]-1\right\} ,
\end{equation}
the additional potential $V$ can be easily introduced in the hybrid Monte
Carlo update procedure of the hybrid-R algorithm \cite{Gottlieb:mq}.
After calculating
the equation of motion for the link variables $U_\mu(y)$,
we find for the
gauge part of the force
\begin{equation}
i\dot{H}_\mu(y)=\left[\frac{\beta}{3}U_\mu(y)T_\mu(y)
\left(1+\frac{\gamma(x-P)}{\beta}\right)\right]_{\rm TA}.
\label{force}
\end{equation}
Here the subscript ${\rm TA}$ indicates the traceless anti-Hermitian part of
the matrix.
We see that in each molecular dynamical step the
measurement of the plaquette is required. However, the only modification
in the gauge force is the factor in round brackets.

\section{Simulation details and the strength of the sign problem}
\begin{figure}
\begin{center}
\begin{minipage}{.48\textwidth}
\raisebox{6.0cm}{(a)}\includegraphics[width=6.5cm, height=6.5cm]{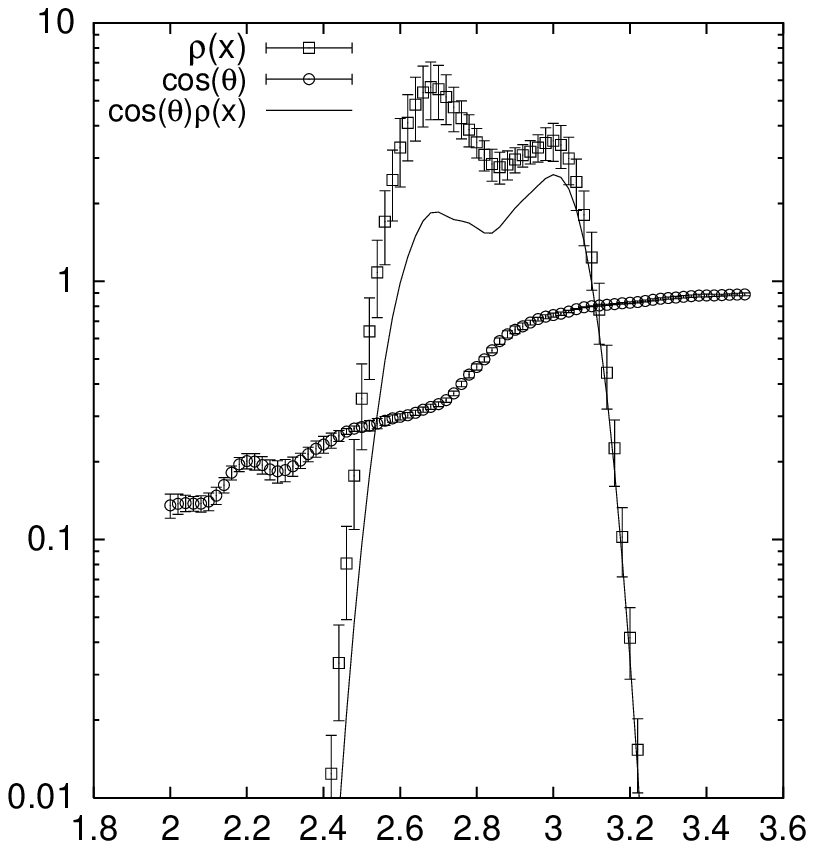}
\end{minipage}
\begin{minipage}{.48\textwidth}
\raisebox{6.0cm}{(b)}\includegraphics[width=7.2cm, height=6.8cm]{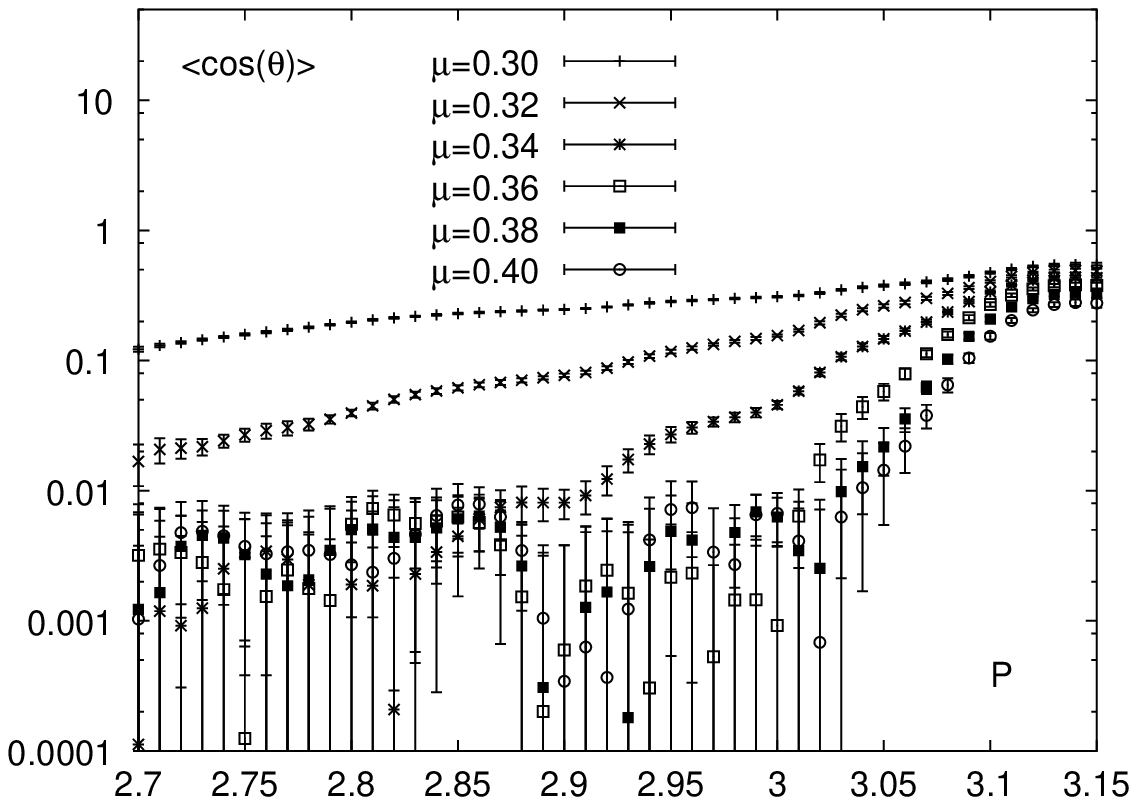}
\end{minipage}
\end{center}
\vspace*{-4mm}
\caption{(a) Results for Simulations at $\beta=4.98$, $\mu=0.3$, $\lambda=0.02$,
  $n_f=4$, $am=0.05$, and number of lattice points: $4^4$. Shown are the density of
  states $\rho(x)$, the phase factor $\left<\cos(\theta)\right>$, and their
  product. (b) Results for Simulations at $\beta=5.1$, $\lambda=0.01$,
  $n_f=4$, $am=0.05$, and number of lattice points: $6^4$. Shown is the suppression
  from the complex phase of the fermion determinant $\left<cos(\theta)\right>$ for 
  different chemical potentials. \label{fig:det_phase}}
\end{figure}

Simulations have been performed with staggered fermions and $N_f=4$. We chose
9 differed points in the $(\beta,\mu)$-plane for the $4^4$ lattice and 8
points for the $6^4$ lattice. On each of these points we did simulations with
20-40 constrained plaquette values, all with quark mass $am=0.05$. Further
simulations has been done with $(\beta,\mu)=(5.1,0.3)$ on the $6^3\times8$
lattice for $am=0.05$ and $am=0.03$.
In order to calculate the plaquette expectation value, or its susceptibility,
one has to perform the following integrals:
\begin{equation}
\left<P\right>=\int dx\; x \rho(x) \left<cos(\theta)\right>_x ,\qquad
\left<P^2\right>=\int dx\; x^2 \rho(x) \left<cos(\theta)\right>_x.
\label{eq:plaq}
\end{equation}
Thus the functions $\rho(x)$ and $\left<cos(\theta)\right>_x$ have to be known
quite precisely. We plot both functions in Figure~\ref{fig:det_phase}(a). 
The transition is signaled in the double peak structure of $\rho(x)$. The phase
factor $\left<cos(\theta)\right>_x$ suppresses the peak of  $\rho(x)$ at smaller
plaquette values, which results in a shift of the critical temperature to
smaller values, in comparison with the phase quenched theory. In
Figure~\ref{fig:det_phase}(b) we show the phase factor for different 
chemical potentials. With increasing chemical potential the phase factor
becomes compatible with zero within errors. In fact, its average value becomes
as low as $cos(\theta)\sim 0.005$. There exist however a small interval
around $P\sim 2.85$, where the phase factor stays finite. In this way, the
Plaquette expectation values is strongly altered by the phase
factor. Figure~\ref{fig:det_phase}(b) 
demonstrates also the advantage of he DOS method over the other approaches of
lattice QCD to finite density. Using the DOS method one is able to do
simulations at directly those Plaquette values which are relevant at finite
density. This solves the so called overlap problem of the reweighting approach.
Furthermore we have checked in \cite{lattice}, that results with have been
obtained within the framework of the DOS method agree very well with earlier
results from the multi-reweighting approach.

\section{The Plaquette expectation value and the phases diagram \label{sec:phases}}
\begin{figure}
\begin{center}
\begin{minipage}{.48\textwidth}
\raisebox{6.0cm}{(a)}\includegraphics[width=6.5cm, height=6.5cm]{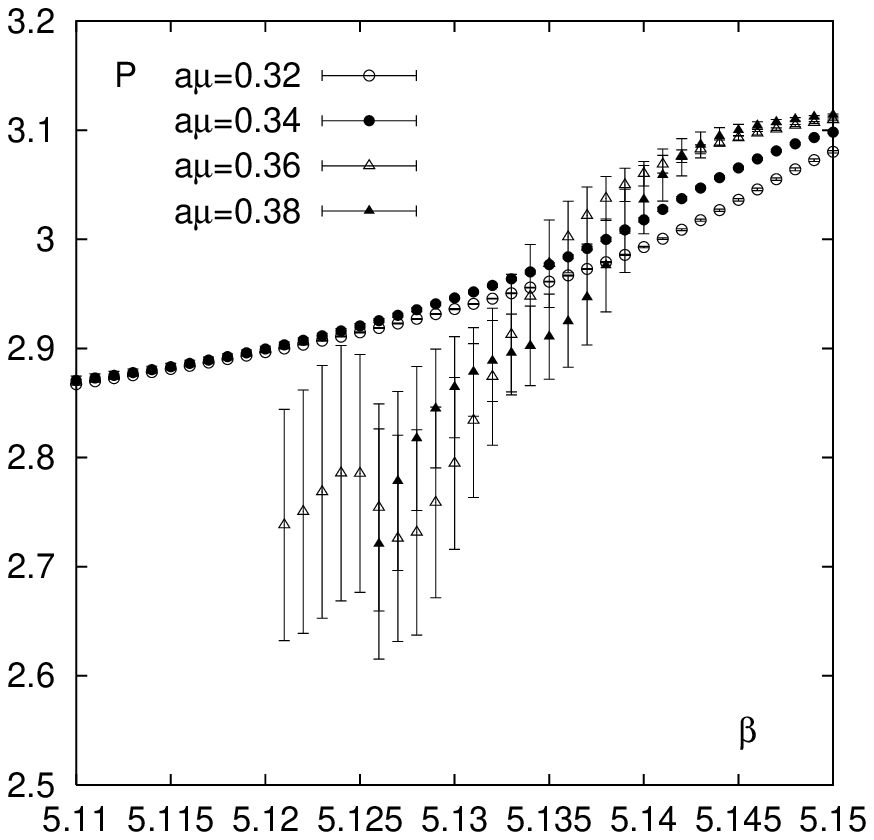}
\end{minipage}
\begin{minipage}{.48\textwidth}
\raisebox{6.0cm}{(b)}\includegraphics[width=6.5cm, height=6.5cm]{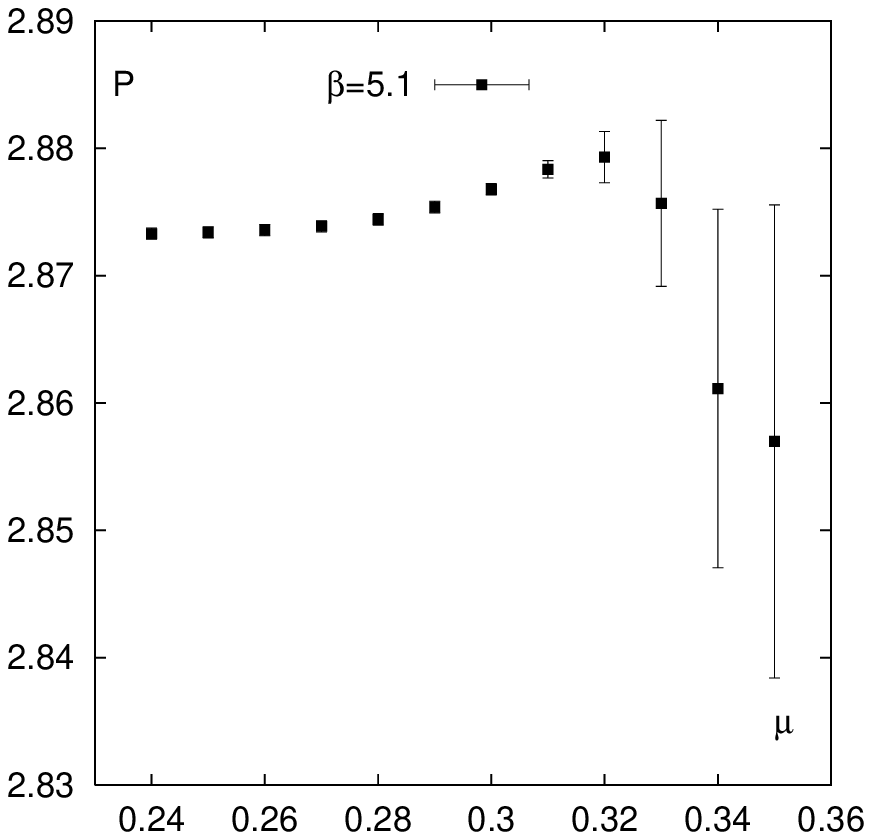}
\end{minipage}
\end{center}
\vspace*{-4mm}
\caption{Results for Simulations at $\beta=5.1$, $\lambda=0.01$,
  $n_f=4$, $am=0.05$, and number of lattice points: $6^4$. Shown is: (a) the
  Plaquette expectation value as a function of the coupling $\beta$ for
  different chemical potentials and (b) the plaquette expectation value at
  fixed coupling, as a function of the chemical potential. \label{fig:plaq}}
\end{figure}

\begin{figure}
\begin{center}
\begin{minipage}{.48\textwidth}
\raisebox{6.0cm}{(a)}\includegraphics[width=6.3cm, height=6.5cm]{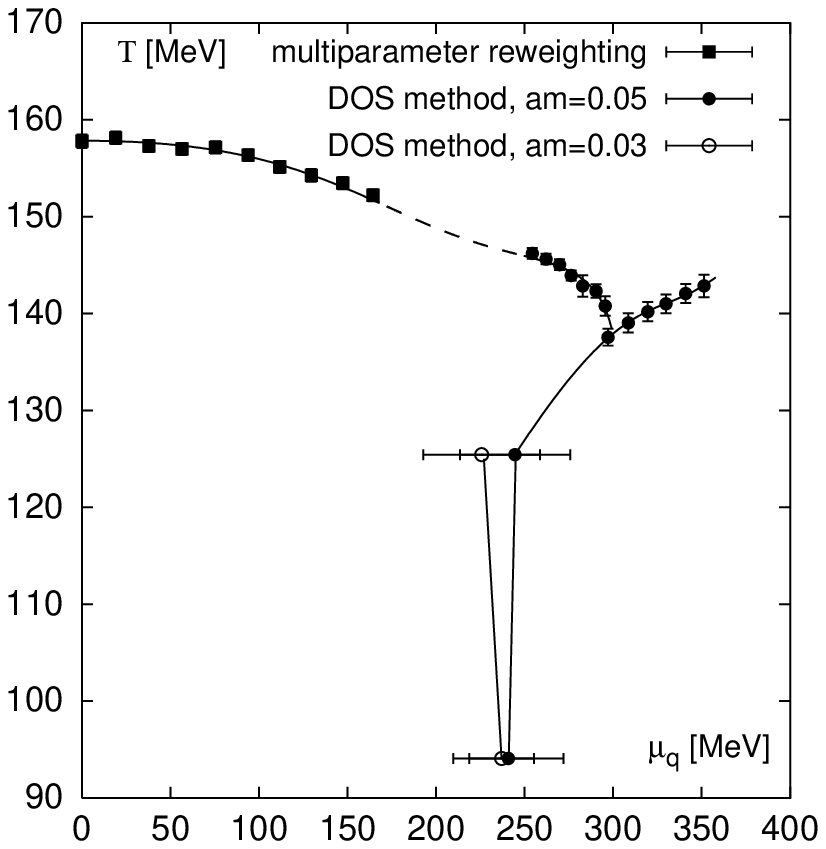}
\end{minipage}
\begin{minipage}{.48\textwidth}
\raisebox{6.0cm}{(b)}\includegraphics[width=6.5cm, height=6.5cm]{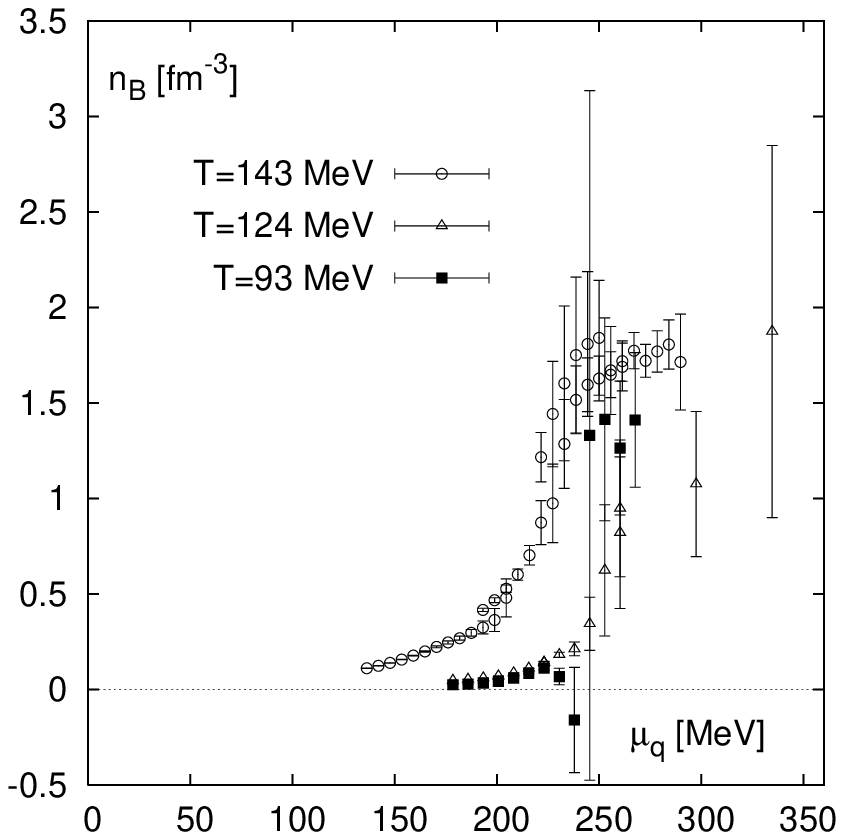}
\end{minipage}
\end{center}
\vspace*{-4mm}
\caption{The phase diagram in physical (a), and the quark number density at
  constant temperature $T=143~\mbox{MeV}$
  ($4^4$ lattice), $T=124~\mbox{MeV}$ ($6^4$ lattice) and $T=93~\mbox{MeV}$
  ($6^3\times 8$ lattice).
\label{fig:phase_diagram}}
\end{figure}

Performing the integration in Eq.~(\ref{eq:plaq}) numerically, we calculate
the plaquette expectation values as shown in Fig.~\ref{fig:plaq}. 
At chemical potentials $\mu\lsim 0.36$, the plaquette signals the QCD
transition through a rapid crossover from a low temperature phase of 
$<P>\sim 2.9$ 
to a high temperature phase of $<P>\sim 3.1$. For $\mu\gsim 0.36$ the plaquette
expectation value at small temperatures drops to $<P>\sim 2.85$. This new low
temperature phase of the plaquette at high   
chemical potentials is caused by the fermion determinant. As on can see in
Figure~\ref{fig:det_phase}(b) the region around $P\sim 2.85$ is the region
which is less suppressed by the phase factor. Another interesting observation
is that the critical coupling, which is decreasing in $\mu$ for $\mu<0.36$
starts to increase for $\mu>0.36$. The plaquette expectation value thus suggests
the existence of three different phases in the ($T$,$\mu$)-diagram with a
triple point, where all those phases coincide. In
Figure~\ref{fig:phase_diagram}(a) we show the phase diagram in physical units. 
The phase boundaries were determined by calculating the peaks in the plaquette
susceptibility. Note, that we make no statement about the order of the
transition lines. To determine the order of the transition one has to perform
finite a finite-size-scaling analysis. 

The scale was set by the Sommer radius $r_0$, measured on a
$10^3\times 20$ lattice. The triple point is located around $\mu_q^{\rm
  tri}\approx 300~\mbox{MeV}$, however its temperature ($T^{\rm tri}$)
decreases from $T^{\rm tri}\approx 148\mbox{MeV}$ on the $4^4$ lattice to
$T^{\rm tri}\approx 137\mbox{MeV}$ on the $6^4$ lattice. This shift reflects
the relatively large cut-off effects one faces, with standard staggered fermions
and temporal extents of 4 and 6.

Also shown in Figure~\ref{fig:phase_diagram}(a) are points from
simulations with quark mass $am=0.03$. The phase boundary turned out to be
--- within our statistical uncertainties --- independent of the the mass.

\section{The quark number density \label{sec:thermodynamics}}
To reveal the properties of the new phase located in the lower right corner of
the phase diagram, we calculated the quark number density, at constant coupling
$\beta$ and at constant temperature respectively. To obtain the density $n_q$
we perform the following integration
\begin{equation}
\label{eq:ddmu}
\left<\frac{{\rm d} \ln {\rm det}M}{{\rm d}(a\mu)}\right>
=\int dx\; \left<\frac{{\rm d} \ln {\rm det}M}{{\rm
      d}(a\mu)}cos(\theta)\right>_x \rho(x)
\label{eq:dmusq}
\end{equation}
The thermodynamic quantity $n_q$ are given as usual by
\begin{equation}
n_q =
\frac{1}{a^3 N_s^3 N_t}
\left<\frac{{\rm d} \ln {\rm det}M}{{\rm d}(a\mu)}\right>
\end{equation}
In Figure~\ref{fig:phase_diagram}(b) we show the baryon number density, which is related
to the quark number density by $n_B=n_q/3$.
The results are plotted in
physical units and correspond to a constant temperature of $T\approx
143~\mbox{MeV}$ ($4^4$ lattice), $T\approx 124~\mbox{MeV}$ ($6^4$ lattice) and
$T\approx 93~\mbox{MeV}$ ($6^4\times 8$ lattice). In order to divide out the leading
order cut-off effect,
we multiply we have multiplied the data with the factor $c=SB(N_t)/SB$, which
is the Stefan-Boltzmann value of a free lattice gas of quarks at a given value of $N_t$,
divided by its continuum Stefan-Boltzmann value. At the same value of the
chemical potential where we find also a peak in the susceptibility of the
plaquette ($\mu_c)$, we see a sudden rise in the baryon number density. Thus
for $\mu>\mu_c$ we enter a phase of dense matter. The transition occurs at a
density of $(2-3)\times n_N$, where $n_N$ denotes nuclear matter density. Above
the transition, the density reaches values of $(10-20)\times n_N$. Quite
similar results have been obtained recently by simulations in the
canonical ensemble \cite{canonical}.


\begin{thebibliography}{99}
\bibitem{Fodor:2001au}
Z.~Fodor and S.~D.~Katz,
\emph{Phys. Lett.}  {\bf B534} (2002) 87
[{\tt hep-lat/0104001}].
%
\bibitem{methods}
Z.~Fodor and S.~D.~Katz,
\emph{JHEP} {\bf 0203} (2002) 014;
%
C.~R.~Allton {\it et al.},
\emph{Phys. Rev.} {\bf D66} (2002) 074507;
R.~V.~Gavai and S.~Gupta,
\emph{Phys. Rev.} {\bf D 68} (2003) 034506;
%
P.~R.~Crompton, [{\tt hep-lat/0301001}];
%
M.~D'Elia and M.~P.~Lombardo,
\emph{Phys. Rev.} {\bf D67} (2003) 014505;
%
P.~de~Forcrand and O.~Philipsen,
\emph{Nucl. Phys.} {\bf B642} (2002) 290;
%
\emph{Nucl. Phys.} {\bf B673} (2003) 170;
%
V.~Azcotiti, {\it et al.}, [{\tt hep-lat/0503010}].
%
\bibitem{overview}
O.~Philipsen,
\emph{PoS} {\bf LAT2005} (2005) 016
[{\tt hep-lat/0510077}].
%
\bibitem{dos}
G.~Bhanot, K.~Bitar and R.~Salvador,
\emph{Phys. Lett.} {\bf B187} (1987) 381;
\emph{Phys. Lett.} {\bf B188} (1987) 246;
M.~Karliner, S.R.~Sharpe and Y.F.~Chang,
\emph{Nucl. Phys.} {\bf B302} (1988) 204;
%
V.~Azcoiti, G.~di~Carlo and A.~F.~Grillo,
\emph{Phys. Rev. Lett.} {\bf 65} (1990) 2239;
%
A.~Gocksch,
\emph{Phys. Rev. Lett.} {\bf 61} (1988) 2054.
%
\bibitem{LUO}
X.~Q.~Luo,
\emph{Mod. Phys. Lett.} {\bf  A16} (2001) 1615.
%
\bibitem{Ambjorn}
J.~Ambjorn, K.~N.~Anagnostopoulos, J.~Nishimura and J.~J.~M.~Verbaarschot,
\emph{JHEP} {\bf 0210} (2002) 062.
%
\bibitem{Gottlieb:mq}
S.~Gottlieb, W.~Liu, D.~Toussaint, R.~L.~Renken and R.~L.~Sugar,
\emph{Phys. Rev.}  {\bf B35} (1987) 2531.
%
\bibitem{lattice}
C.~Schmidt, Z.~Fodor and S.~Katz,
\emph{PoS} {\bf LAT2005} (2005) 163
[{\tt hep-lat/0510087}].
%
\bibitem{canonical}
A.~Alexandru, M.~Faber, I.~Horvath and K.~F.~Liu,
[{\tt hep-lat/0410002}];
S.~Kratochvila and P.~de~Forcrand,
\emph{PoS} {\bf LAT2005} (2005) 167
[{\tt hep-lat/0509143}].
%
\end{thebibliography}
\end{document}